# Speed of sound in methane under conditions of planetary interiors

Thomas G. White,[1,2,*] Hannah Poole,[2] Emma E. McBride,[3,4] Matthew Oliver,[1,2,5] Adrien Descamps,[3,4]
Luke B. Fletcher,[3] W. Alex Angermeier,[1] Cameron H. Allen,[1] Karen Appel,[6] Florian P. Condamine,[7] Chandra B. Curry,[3,8]
Francesco Dallari,[9] Stefan Funk,[10] Eric Galtier,[3] Eliseo J. Gamboa,[3] Maxence Gauthier,[3] Peter Graham,[11] Sebastian Goede,[6]
Daniel Haden,[1] Jongjin B. Kim,[3] Hae Ja Lee,[3] Benjamin K. Ofori-Okai,[3] Scott Richardson,[11] Alex Rigby,[2]
Christopher Schoenwaelder,[3] Peihao Sun,[9] Bastian L. Witte,[3] Thomas Tschentscher,[6] Ulf Zastrau,[6] Bob Nagler,[3]
J. B. Hastings,[3] Giulio Monaco,[9] Dirk O. Gericke,[12] Siegfried H. Glenzer,[3] and Gianluca Gregori[2]

[1]*Department of Physics, University of Nevada, Reno, Nevada 89557, USA*
[2]*Department of Physics, University of Oxford, Parks Road, Oxford OX1 3PU, United Kingdom*
[3]*SLAC National Accelerator Laboratory, 2575 Sand Hill Road, Menlo Park, California 94025, USA*
[4]*School of Mathematics and Physics, Queens University Belfast, University Road, Belfast BT7 1NN, United Kingdom*
[5]*Central Laser Facility, STFC Rutherford-Appleton Laboratory, Chilton OX11 0QX, United Kingdom*
[6]*European XFEL, Holzkoppel 4, 22869 Schenefeld, Germany*
[7]*Extreme Light Infrastructure ERIC, ELI Beamlines Facility, Za Radnici 835, CZ-252 41 Dolni Brezany, Czech Republic*
[8]*Department of Electrical and Computer Engineering, University of Alberta, Edmonton, Alberta, Canada T6G 1H9*
[9]*Dipartimento di Fisica e Astronomia "Galileo Galilei", Università degli Studi di Padova, Via F. Marzolo, 8, I-35131 Padova, Italy*
[10]*Friedrich-Alexander-Universität Erlangen-Nürnberg, Erlangen Centre for Astroparticle Physics, Nikolaus-Fiebiger-Str. 2,
91058 Erlangen, Germany*
[11]*AWE, Aldermaston, Reading, Berkshire RG7 4PR, United Kingdom*
[12]*Centre for Fusion, Space and Astrophysics, Department of Physics, University of Warwick, Coventry CV4 7AL, United Kingdom*

We present direct observations of acoustic waves in warm dense matter. We analyze wave-number- and energy-resolved x-ray spectra taken from warm dense methane created by laser heating a cryogenic liquid jet. X-ray diffraction and inelastic free-electron scattering yield sample conditions of $0.3 \pm 0.1$ eV and $0.8 \pm 0.1$ g/cm$^{-3}$, corresponding to a pressure of ∼13 GPa. Inelastic x-ray scattering was used to observe the collective oscillations of the ions. With a highly improved energy resolution of ∼50 meV, we could clearly distinguish the Brillouin peaks from the quasielastic Rayleigh feature. Data at different wave numbers were utilized to derive a sound speed of $5.9 \pm 0.5$ km/s, marking a high-temperature data point for methane and demonstrating consistency with Birch's law in this parameter regime.

DOI: 10.1103/PhysRevResearch.6.L022029

The state and evolution of methane-rich planets, such as Uranus and Neptune, are primarily determined by the properties of dense, compressed matter within their interiors [1–4]. Models rely on accurate knowledge of thermodynamic and transport properties [5], however, a complex interplay of factors, including partial ionization, electron quantum degeneracy, and moderate to strong ion-ion interactions, renders theoretical predictions challenging [6]. This complexity is particularly pronounced in carbon-bearing materials, where intriguing high-pressure chemistry unfolds [3,7,8]. While static material properties are well constrained through a combination of experiments [3,7,9] and *ab initio* simulations [10–12], significant uncertainties persist in understanding dynamic and transport properties [13,14], especially in the 10–600 GPa and 1000–15 000 K range [15].

One vital property is the sound speed ($c_s$), which plays a crucial role in determining the adiabatic exponent [16], bulk modulus [17], and thermal conductivity [18,19]. At low temperatures, methane (CH$_4$) has been observed to conform to Birch's law [20,21], where the sound speed shows a linear dependence on material density ($\rho$),

$$c_s \text{ (km/s)} = M\rho \text{ (g/cm}^3\text{)} + B. \quad (1)$$

This empirical relationship often serves as a basis for extrapolating sound speeds to conditions within planetary interiors, with the coefficients $M = 13.2$ and $B = -4.7$ established through room-temperature measurements [21]. However, the linear behavior of Eq. (1) and its coefficients remain untested at the high temperatures found in planetary interiors. Indeed, in warm dense iron, the coefficients were found to vary considerably with temperature [17,22,23].

In this Letter, we describe the development and application of high-resolution inelastic x-ray scattering to directly observe acoustic waves in methane at elevated temperatures. X-ray Thomson scattering (XRTS) utilizing high-intensity beams of penetrating radiation has emerged as a way to determine the microstructure and dynamics of extreme states of matter [24–28]. However, experiments have previously been limited to bandwidths of tens of eV for laser-produced x rays [24] down to a few eV when using the seeded mode of x-ray free-electron lasers [27], whereas the energy shift caused by the collective ion motion is at a scale of just a few tens of meV. Combining improvements in free-electron laser technology with a silicon monochromator developed at synchrotron light sources, we were able to generate x rays with sufficiently narrow bandwidths to distinguish the Brillouin peaks from the elastic Rayleigh peak. In our experiment, we attained a 50 meV energy resolution [29]. On this basis, we have directly determined the sound speed in a warm dense matter sample by analyzing the dispersion of the inelastic ion modes, demonstrating consistency with Birch's law at a temperature of 0.3 eV.

---

*Corresponding author: tgwhite@unr.edu

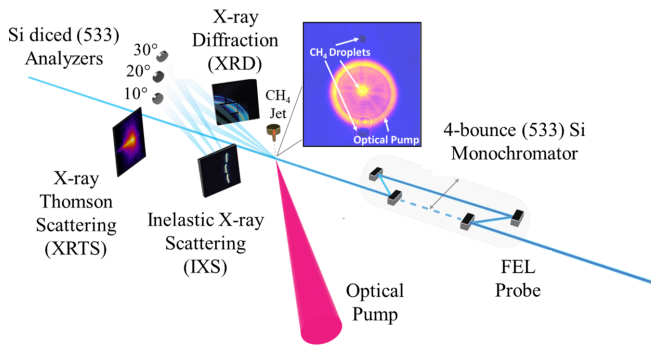

FIG. 1. Setup of the main diagnostics. The x-ray beam is scattered off a fs-irradiated 5-μm-diameter cryogenic liquid methane jet at 5 Hz repetition rate. With the monochromator applied, the energy of the inelastic scattered photons is determined using Si diced crystal analyzers at scattering angles of 10° ($Q = 0.66$ Å$^{-1}$), 20° ($Q = 1.32$ Å$^{-1}$), and 30° ($Q = 1.96$ Å$^{-1}$). Without the monochromator, the angular spread of the photons (XRD) is captured on Cornell SLAC pixel array detectors covering approximately 10°–50° (0.5–3.1 Å$^{-1}$), while the inelastic component (XRTS) is measured using a HAPG crystal positioned at a scattering angle of 47° ($Q = 3.03$ Å$^{-1}$). The blue panel depicts the sample illuminated by the optical pump, illustrating the substantial size difference between the laser spot and methane droplets.

Figure 1 shows a schematic of the experiment performed at the Matter in Extreme Conditions (MEC) endstation at the Linac Coherent Light Source (LCLS). The incident x-ray pulse, with a central energy of 7492.1 eV and a 40 fs duration, was focused to a ∼5-μm spot on the sample using a stack of beryllium compound refractive lenses. When measuring the collective motion of ions, a four-bounce silicon channel-cut (533) monochromator at a Bragg angle of 87.7° was driven into the beam to reduce the bandwidth to a nominal ∼32 meV. The seeded mode of LCLS was used to maximize the throughput of the monochromator. This mode has a bandwidth of ∼1 eV [compared to ∼20 eV in the self-amplified spontaneous emission (SASE) mode], which significantly increased the fluence through the monochromator [30].

Three diced crystal analyzers [31] were positioned vertically at angles of 10°, 20°, and 30°, corresponding to scattering vectors $Q$ of 0.66, 1.32, and 1.96 Å$^{-1}$, respectively. The analyzers are made up of thousands of 1.5 mm × 1.5 mm flat silicon crystals, which form a 10-cm-diameter spherically curved crystal. The spectrometer used energy-dispersive Johann geometry and was based on a Rowland circle with a diameter of 1 m and a Bragg angle matched to that of the monochromator. The spectrum measured by each individual flat crystal is combined on the detector, resulting in an increased signal. The overall energy resolution of the monochromator and diced crystal analyzer is ∼50 meV over a range of ∼500 meV. The scattered x rays were collected on an ePIX100 detector with a 50 μm pixel size [32]. We anticipate approximately $10^{10}$ photons per x-ray pulse reaching the sample, resulting in a beam intensity of $7.6 \times 10^{14}$ W/cm$^2$. A detailed description of the spectrometer setup can be found in Ref. [29].

The sample consisted of cryogenic liquid methane droplets [33,34] that were illuminated with the uncompressed short pulse Ti:sapphire laser with an energy of 600 mJ at a wavelength of 800 nm. The spot was spatially and temporally Gaussian with a full width at half maximum of 100 μm and 150 ps, respectively. The optical and x-ray laser pulses were overlapped 5 mm from the end of the jet nozzle to prevent potential damage. The replenishing jet target allowed the experiment to be run at a repetition rate of 5 Hz, allowing for considerably improved photon statistics compared to single-shot experiments.

Two ancillary diagnostics, XRTS with ∼20 eV energy resolution and x-ray diffraction (XRD), characterized the thermodynamic state of the system. Both diagnostics were run without the monochromator in place to maximize signal levels. The XRD employed a Cornell Stanford pixelated area detector (CSPAD) placed approximately 100 mm from the interaction point covering a 2θ range of 10°–50° (0.5–3.1 Å$^{-1}$). Processing and integration of the powder patterns were carried out using the DIOPTAS software [35], and the diagnostic was calibrated using x-ray diffraction from polycrystalline LaB$_6$ [36].

Figure 2 shows the evolution of the angular-resolved scattering as a function of laser delay. When no optical laser is present ("X-ray Only"), we see three distinct Bragg peaks. These are consistent with methane forming face-centered-cubic (fcc) crystallites as the liquid jet freezes when exiting the nozzle. From the Bragg peak positions, we infer a density of 0.5 g/cm$^3$. At 0 ps, which corresponds to the x-ray pulse arriving at the peak of the drive laser, we observe significant heating and the coexistence of the solid and liquid phases, with the intensity of the Bragg peaks decreasing and a broad ion-ion liquid correlation peak emerging. At an additional delay of 150 ps, the signal is dominated by a broad liquid correlation peak.

The XRTS spectrum was collected on a cylindrically curved, highly annealed pyrolytic graphite (HAPG) crystal in von Hamos geometry. The spectrometer was positioned at an average scattering angle of 47°, with an acceptance angle of 7°. The spectrum was captured on an ePix100 detector with a 50 μm pixel size. The energy dispersion was found to be 1.93 eV/px, calibrated using fluorescence of the $K\beta_1$ line of iron [37]. Figure 3 shows the spectrum collected at a time delay of 150 ps, coinciding with the XRD signal shown in the bottom panel of Fig. 2. In addition to a reduction in inelastic scattering due to the heating and compression of the methane, the modified spectrum now contains a significant scattering contribution from a hot blow-off plasma surrounding the irradiated target, as highlighted by the large free-free Compton feature visible in the bottom inset. The top inset of Fig. 3 shows the scattered spectrum for undriven methane, which is consistent with a synthetic spectrum [38–42] calculated for a density of 0.5 g/cm$^3$ at 0.025 eV, confirming the density inferred from the XRD data.

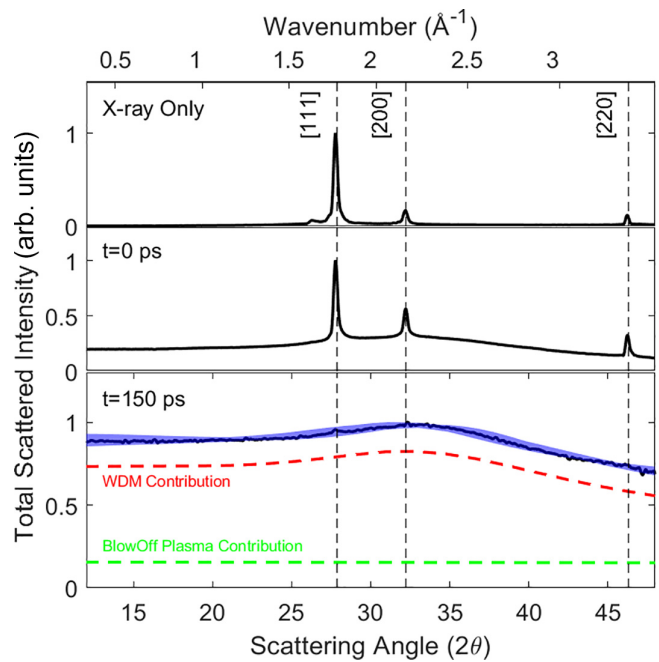

FIG. 2. Angular-resolved x-ray diffraction (XRD) data. Top: Bragg peaks from Debye-Scherrer rings indicative of an fcc lattice of methane at a density of 0.5 g/cm$^3$ are observed. Middle: During the laser pulse, we observe the emergence of a broad liquid diffraction peak concurrently with Bragg peaks of reduced intensity. Bottom: The Bragg peaks have almost disappeared, and the signal is dominated by liquid diffraction. The shaded region shows the range of accepted fits corresponding to thermodynamic conditions given in Table I. The red and green curves show the relative contribution from the high-density matter and blow-off plasma.

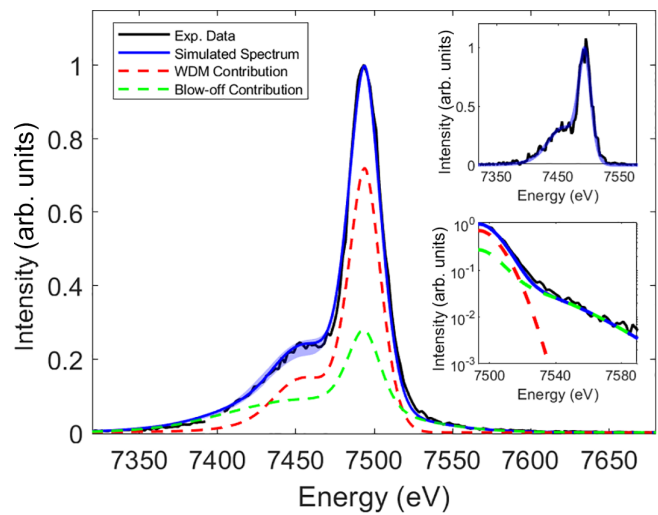

FIG. 3. X-ray Thomson scattering spectra at 47° for methane 150 ps after the laser peak. The experimental data (black line) are fitted with theoretical predictions considering scattering from two regions with distinct thermodynamic conditions: warm dense methane (red) and a hot, low-density blow-off plasma (green). The blue-shaded region highlights the range of spectra accepted by the fitting procedure. The top inset shows the undriven case, which has a higher noise level due to a lower scattering efficiency into the detector, which lies outside the largest Bragg peaks. The bottom inset highlights the high-energy wing of the Compton feature, which is used to constrain the temperature of the blow-off plasma [36,43].

A method for exploring the complex multiparameter space is to utilize Bayesian inference using Markov-chain Monte Carlo (MCMC) [44,45]. Given a specific set of parameters, $\Theta(T_e, \rho, Z_H, Z_C, r_c)$, the MCMC process calculates the likelihood of these sets producing the given experimental spectrum and provides a robust estimate of the errors. Two separate MCMC explorations were run for the XRTS and XRD data. Each assumed two uniform plasma conditions, one representing the warm dense methane in the center of the target and the other the blow-off plasma that directly interacted with the heating laser. The blow-off plasma was found to be at $T_e = $ 36 eV and assumed to be at a low density, $\rho < 0.05$ g/$cm^3$. At these conditions, the ionization of the two species is estimated to be $Z_C = 4$, $Z_H = 1$ [46]. The regions were weighted by the parameter $r_c$, which is equal to the fraction of scattering from warm dense methane determined to contribute to the overall signal. Additional details on the MCMC procedure are provided in the Supplemental Material [36].

The range of accepted spectra is shown as blue-shaded regions in Figs. 2 and 3 for the XRD and XRTS diagnostics, respectively. Table I shows the inferred thermodynamic properties of the compressed, warm methane after combining the information from XRTS and XRD. The error on each parameter is quoted as the 1$\sigma$ deviation from the mean. The values of density and temperature with the highest likelihood are 0.3 eV (3480 K) and 0.8 g cm$^{-3}$, corresponding to a pressure of ∼13 GPa found from *ab initio* density functional theory [10,15]. Under these conditions, the system is expected to be on the cusp between the molecular, plasma, and diamond formation states [10,47]. However, our XRD data reveal no indication of diamond formation or molecular liquid methane, which typically exhibits a shoulder on the low-$k$ side of the liquid peak [48].

TABLE I. Thermodynamic conditions of the warm dense methane found in the experiment. The experimental error bars are obtained from the MCMC fitting procedure [36]. For comparison, the conditions at the proposed thermal boundary layer (TBL) of Uranus are shown.

| Parameter | Experiment | TBL [15,49,50] |
|---|---|---|
| $T_e$ (eV) | 0.3 ± 0.1 | 0.2–0.6 |
| $\rho$ (g/cm$^3$) | 0.8 ± 0.1 | 0.3–1.5 |
| $P$ (GPa) | 13 ± 1 [10,15] | 10–15 |
| $Z_C$ | 0.4 ± 0.1 | |
| $Z_H$ | 0.4 ± 0.1 | |
| $r_c$ | 0.73 ± 0.01 | |

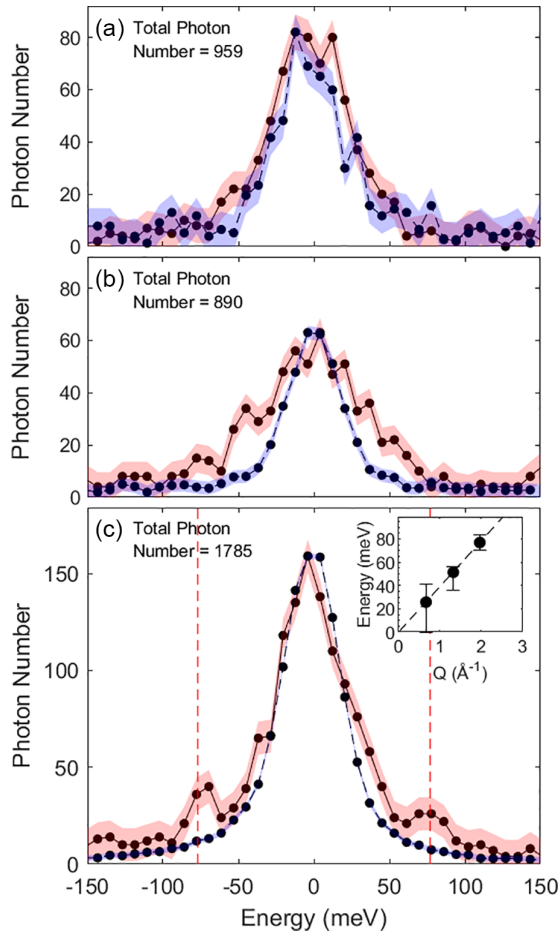

FIG. 4. High-resolution inelastic x-ray scattering spectra taken at (a) 10°, (b) 20°, and (c) 30° for driven (red) and undriven (blue) methane. The shaded areas indicate the Poisson noise limit, which is equal to $\sqrt{N}/N$. The data consists of 5965 laser shots with the number of photons collected in each case shown. The data are presented in bins with a width of 8.2 meV, corresponding to a single pixel on the detector. The vertical dashed lines show the predicted peak position for an acoustic wave with a sound speed of 5.9 km/s. The inset shows the dispersion relation obtained by fitting each spectrum with a combination of three Voigt profiles [36], with the dashed line corresponding to the measured acoustic wave speed.

The high-resolution inelastic x-ray scattering spectra are shown in Fig. 4. The three detectors are situated in the collective scattering regime [39], exhibiting scattering parameters ($\alpha = 1/Q\lambda_s$) of 13.8, 6.9, and 4.6. These values were computed using a Debye screening length of $\lambda_s = 0.11$ Å, utilizing the thermodynamic parameters in Table I. Using scattering from undriven methane, shown by the blue lines in Fig. 4, the instrument function for the analyzers was found to have a full width at half maximum (FWHM) of 49, 42, and 46 meV for 10°, 20°, and 30°, respectively. These values are consistent with the calculated value of 46 meV [29].

The driven case is shown by the red lines in Fig. 4. On the detector at 30°, which corresponds to the highest scattering vector, Brillouin peaks were observed at an energy transfer of around 75 meV on either side of the Rayleigh peak. Furthermore, as a consequence of detailed balance, the height of the downshifted peak is larger than the upshifted peak [51]. The intensity ratio between the two peaks is $0.8 \pm 0.1$, corresponding to an *ion* temperature of roughly 0.3 eV, in agreement with the *electron* temperature measured through XRTS and XRD, which is better constrained. This alignment serves to confirm that the system is in local thermodynamic equilibrium. On the detector positioned at the 20° scattering angle, the Brillouin peaks are not well resolved, instead appearing as additional shoulders on either side of the Rayleigh peak. Finally, no additional broadening was observed on the detector positioned at 10°.

In the viscoelastic regime, the shape of the scattering spectrum can be described in terms of $Q$- and $\omega$-dependent transport coefficients [52,53]. In our case, the low photon number and consequential energy resolution preclude accurate extraction of such quantities directly from the shape and width of the spectrum. Nevertheless, a more robust measurement is the determination of the acoustic wave speed from the Brillouin peak positions. In the hydrodynamic regime ($\alpha < 2.32$ or, equivalently, $Q < 0.43/\lambda_s = 3.9$ Å$^{-1}$), these positions exhibit a linear shift with wave number ($\omega = E/\hbar = c_s Q$) with $c_s$ being the speed of acoustic waves [54]. In our case, the peak position on the 30° detector matches an acoustic wave speed of $5.9 \pm 0.5$ km/s, as shown by the red vertical dashed lines in Fig. 4. The inset of Fig. 4(c) shows the dispersion relation obtained by fitting each of the driven spectra with a combination of three Voigt profiles [36].

Our realization of high-resolution x-ray scattering has enabled us to resolve the acoustic waves and measure the sound speed at conditions comparable to the extreme conditions found in planetary interiors. Our sound speed measurement is consistent with Eq. (1), with coefficients derived at room temperature, which for a density of 0.8 g/cm$^3$ predicts a sound speed of 5.86 km/s. Our findings suggest little variation in the sound speed of methane from room temperature up to 0.3 eV (3480 K), confirming previous predictions that temperature effects are limited for these excitations [21]. However, data points at additional thermodynamic conditions are required to conclusively verify the linear behavior of Birch's law at elevated temperatures.

This measurement represents a reliable benchmark for equation-of-state (EOS) models applied in this region, surpassing the limitations imposed by solely static thermodynamic properties or the microscopic structure [55]. Testing the accuracy of EOS models is of immense importance in advancing comprehensive descriptions of the formation, thermal evolution, and internal structure of the ice giants within our solar system and a large class of exoplanets. For example, the conditions achieved in our experiment closely resemble those predicted for the proposed thermal boundary layer (TBL) of Uranus, shown in Table I. The TBL is expected to occur at a radius of approximately 20 000 km, positioned between the inner and outer envelopes, and is considered one possible solution to explain the observed low luminosity of Uranus [15,49,50]. The sound speeds in this regime also play a pivotal role in the field of giant planet seismology, enabling us to employ Uranus quakes and ring seismology as tools to unravel the planet's internal composition [16,56,57].

In summary, we have demonstrated an x-ray scattering platform capable of assessing elemental properties, here sound speeds, within high-pressure and high-temperature states akin to those found in planets of our solar system and exoplanets. Notably, we have provided a high-temperature data point for methane and have shown that it is consistent with Birch's law using the same coefficients as obtained for room temperature.

T.G.W. thanks Yafis Barlas for valuable discussions. This material is partially based upon work supported by the U.S. Department of Energy, Office of Science, Office of Fusion Energy Sciences (Award No. DE-SC0019268) and the National Nuclear Security Administration (NNSA) (Award No. DE-NA0004039). This work was supported by the U.K. Research & Innovation Future Leaders Fellowship (MR/W008211/1) and by the Science and Technologies Facilities Council under Grant No. ST/W000903/1. K.A. and T.T. acknowledge support from the DFG (Grant No. FOR 2440). The work of E.E.M., A.D., L.B.F., F.P.C., C.B.C., E.J.G., M.G., S.G., J.B.K., B.K.O.-O., C.S., B.L.W., D.O.G., and S.H.G. were supported by the U.S. DOE Office of Science, Fusion Energy Sciences under FWP No. 100182. C.B.C. acknowledges partial support from the Natural Sciences and Engineering Research Council of Canada (NSERC). Use of the Linac Coherent Light Source, SLAC National Accelerator Laboratory, is supported by the U.S. Department of Energy, Office of Science, Office of Basic Energy Sciences under Contract No. DE-AC02-76SF00515. The MEC instrument is supported by the U.S. Department of Energy, Office of Science, Office of Fusion Energy Sciences under Contract No. DE-AC02-76SF00515.